# Vector Measurements Using All Optical Scalar Atomic Magnetometers

Rui Zhang, [1,a)] Rahul Mhaskar, [2)] Ken Smith, [1)] Easswar Balasubramaniam, [1)] and Mark Prouty [1)]

[1]*Geometrics, Inc., San Jose, CA, 95131, USA*

[2] *formerly Geometrics, Inc., San Jose, CA, 95131, USA, now Keysight Technologies, Santa Clara, CA, 95051, USA*

Vector field measurement is demonstrated with an all-optical scalar atomic magnetometer using intrinsic parameters related to its scalar operation. The Bell-Bloom type atomic magnetometer measures the Larmor precession of cesium atoms through on-resonant absorption of a probe beam. While the AC component of the probe signal is used for the field magnitude measurement, the probe DC signal contains information about the polar angle, defined as the angle between the magnetic field and the probe beam. Additional polar angle information is obtained from the light-shift-induced magnetic field caused by the frequency modulation of the probe beam. With a measurement time of 100 milliseconds, better than 0.02° sensitivity has been achieved using a commercial miniaturized sensor at the optimal sensor orientation. The angle measurement accuracy is checked against an optical encoder over the entire polar angle range of 0º to 180º. Better than 1º error is recorded over most set polar angles. Azimuthal angle measurement is also exhibited with two orthogonally oriented sensors.

## I. INTRODUCTION

Scalar atomic magnetometers measure the Larmor precession frequency of atomic spins, which depends on the magnitude of the background magnetic field rather than its direction. Due to this unique property these devices have been widely used in geophysics applications ranging from archaeological imaging to mineral exploration and unexploded ordnance (UXO) detection. All these applications require mobile platforms where orientation-induced noise is a big issue for vector magnetometers. Scalar atomic magnetometers can be found in other applications [1], such as biomagnetic field sensing in an unshielded environment [2], because of their supreme noise performance and operation in finite magnetic fields. Better than 300 fT/√Hz sensitivities have been achieved in miniaturized vapor cells [3, 4, 5, 6] and with bigger cells much better sensitivities have been demonstrated [7, 8, 9, 10, 11, 12, 13, 14]. However, in some applications, such as magnetic navigation and real-time UXO localization [15, 16], it is important to know the magnetic field direction as well. Acquiring vector measurement capability will certainly help scalar atomic magnetometers be useful in more applications since such a device will provide the complete knowledge of the magnetic field.

Many vector measurement schemes have been explored using atomic magnetometers. For example, vector field information can be derived from the bias magnetic fields applied to atoms to achieve a zero-field condition [17, 18]. Researchers have also realized vector magnetometers based on the nonlinear magneto-optical rotation effect [19] and the Voigt effect [20]. In free spin precession magnetometers, it has been demonstrated that transient oscillation signals can be used to construct the vector fields [21, 22, 23]. It is also possible to apply additional optical beams to atoms either to create effective light-shift induced magnetic fields and compare the precession frequency shift to obtain a measurement of the field angle [24], or to probe the precession in different directions and rely on the relative phase or amplitude of the precession signals to determine the field direction [25]. In this paper, we study a Bell-Bloom [26] type scalar atomic magnetometer employing a pump-probe scheme. Only intrinsic signals associated with the operation of the scalar magnetometer are used to extract the field angle information. Therefore, the performance of the scalar operation is not affected by the vector field measurement. The field angle measurement scheme has been tested using commercially available miniaturized atomic sensors. After the laboratory investigation, the

a) Email: rzhang@geometrics.com

sensors are mounted on a rotation table and tested in an unshielded environment. The angle measurement is compared with the result obtained by an optical encoder embedded in the rotational axis of the table.

## II. METHOD

We use MFAM magnetometers from Geometrics for this experiment. The MFAM sensor employs a pump-probe scheme with the pump beam tuned to the |F=3> → |F′=4> transition of Cs D1 line and the probe on the |F=4> → |F′=3> transition. Both beams are circularly polarized. Detailed information about the sensor can be found in our previous publication [27]. As shown in Figure 1, the sensor is placed inside a 4-layer magnetic shield. The background magnetic field, **B**, is controlled by two pairs of Helmholtz coils in x and z directions. The angle between the probe beam and **B** is given by $\theta$ and referred to as the polar angle. After transmitting through the Cs vapor cell, the probe beam is detected by a photodiode, whose current signal is then converted to voltage by a transimpedance amplifier. The AC component of the probe transmission contains the Larmor frequency of the atomic spin and is processed by the magnetometer driver to yield the magnitude of **B**. The DC part of the signal is normally ignored. However, the probe DC transmission has a strong dependence on the polar angle $\theta$ and can be used as a signal to extract the angle information.

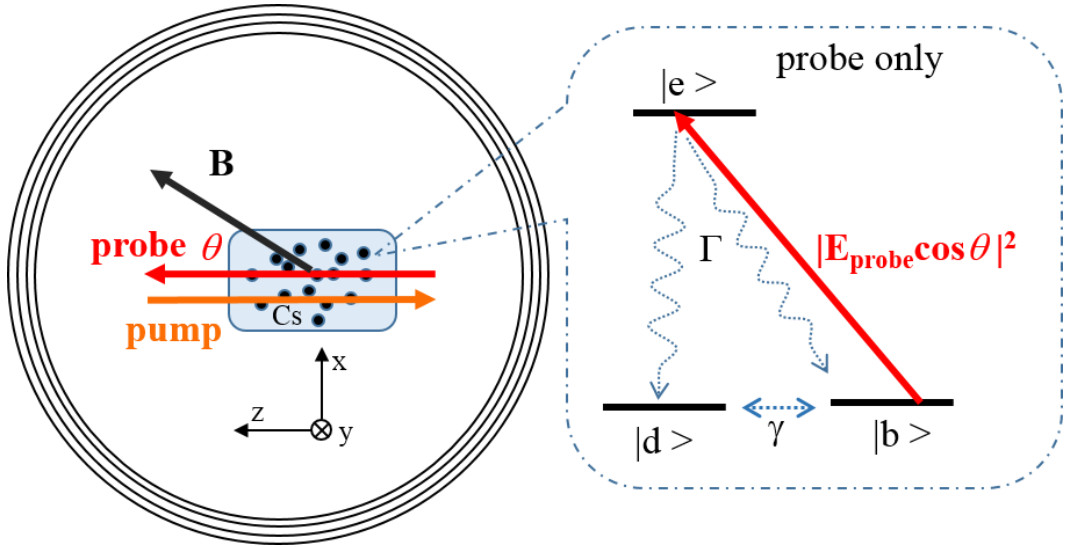

Figure 1 Schematics showing the MFAM sensor inside a 4-layer magnetic shield. The magnetic field, **B**, is controlled by two pairs of Helmholtz coils in x and z directions. The probe-only transition diagram in the basis defined by the magnetic field is also shown.

The dependence of the probe DC transmission on $\theta$ can be explained as follows. To σ- probe light, |F=4, m = -4> and |F=4, m = -3> magnetic sublevels are dark states and atoms in these states do not interact with the probe light. When **B** and the probe beam are parallel, $\theta = 0°$ or $180°$, most atoms will be pumped into the dark state by the probe and the probe DC transmission will be high. However, in the case of other $\theta$ values, **B** will rotate the atoms out of the dark states through the precession and the probe DC transmission will decrease. In a more quantitative analysis, we can refer to the dark states as |d> and the other magnetic sublevels as the bright state |b>. In the atomic basis defined by the magnetic field, the probe Rabi frequency is proportional to $cos^2\theta$. As shown in Figure 1, if the decay rate, Γ, of the excited state |e> is much larger than the probe Rabi frequency and the ground state relaxation rate γ, which is the case here, a simple rate equation calculation can yield that the steady state atomic population in |d> is proportional to $cos^2\theta$. In other words, the atomic population in |b> can be expressed as $A + B * sin^2\theta$. The probe transmission signal, V, should then follow the equation $V = C * \exp[-D * sin^2(\theta)]$. Here $A, B,$

*C* and *D* are constants that depend on the cell temperature and the probe power. The *V* vs *θ* relation is confirmed experimentally. With the pump beam, the transition diagram becomes more complicated. But the effect of the pump beam on the |b> state population can be treated as adding a higher order term so that the probe transmission *V* as a function of *θ* becomes:

$$V = C_0 * \exp[-C_1 * \sin^2(\theta) - C_2 * \sin^4(\theta)] \quad (1)$$

Again $C_0$, $C_1$ and $C_2$ are constants, affected by the cell temperature and the probe and the pump powers, but insensitive to the field magnitude. From Equation (1), it is obvious that with the calibrated constants $C_0$, $C_1$ and $C_2$, *θ* can be calculated if *V* is measured.

Note that according to Equation (1) *V* is symmetric with respect to the plane defined by *θ*=90°. Hence the probe transmission method cannot distinguish between *θ* and 180°-*θ*. We introduce a supplementary angle measurement method to solve this problem. The method relies on the light-shift effect [28]. A circularly polarized probe beam causes a slight shift of energy levels of atomic ground states, equivalent to a fictitious magnetic field being applied along the light propagation direction. The total magnetic field, which is measured by the magnetometer, is given by the vector sum of the real magnetic field ***B*** and the effective magnetic field ***B**_k*. In general, $B_k << B$ so that the total magnetic field can be approximated as $B+B_k\cos\theta$. If we apply a small oscillating current at a frequency *f* to the probe laser, $B_k$ will also develop an oscillating component with the same frequency *f*, $B_k=B_{0k}+B_{1k}*\sin(2\pi ft)$. When the magnetometer output is demodulated at the frequency *f*, the component $B_{1k}\cos\theta$ can be extracted. With the calibrated $B_{1k}$, obtaining the value of $B_{1k}\cos\theta$ yields a measurement of *θ*. This method is essentially the same as described in the reference [24] except without introducing additional laser beams.

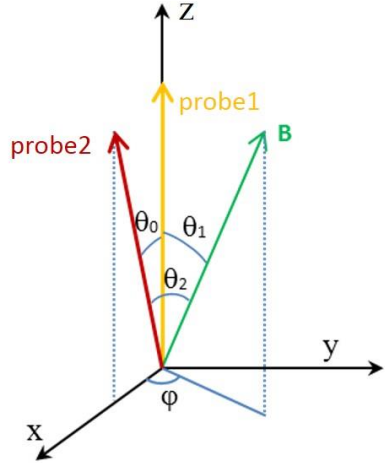

Figure 2 Plot showing the relative directions of two probes and the magnetic field. The azimuthal angle, φ, of the magnetic field can be calculated based on the two polar angles, $\theta_1$ and $\theta_2$, and the angle separation between the two probes, $\theta_0$.

Knowing the polar angle *θ* does not complete the vector measurement. We still need the azimuthal angle φ of the magnetic field. This can be accomplished by combining the polar angle measurements from two sensors. As shown in Figure 2, one probe is aligned along the z-axis and the other is on the x-z plane and off by $\theta_0$. The polar angles, $\theta_1$ and $\theta_2$, with respect to each probe are calculated based on the methods presented in the previous paragraphs. Based on the geometry, the azimuthal angle φ of the magnetic field satisfies the following equation:

$$\cos\varphi = \frac{\cos\theta_2 - \cos\theta_1 \cos\theta_0}{\sin\theta_1 \sin\theta_0} \quad (2)$$

If we align the two sensors such that $\theta_0 = 90°$, Equation (2) can be greatly simplified. The azimuthal angle can then be expressed as $\varphi = arccos(\cos\theta_2/\sin\theta_1)$. Note that *arccosine* is only defined between 0° and 180°. The azimuthal angle is still not distinguishable between φ and 360°- φ. This may not be an issue in most applications since we can always orient the sensors such that the magnetic field is close to the y direction initially. In practice, the mobile platforms usually do not vary the sensor orientation by much. Another solution is to add a low-cost compass or an oscillating field along the y-direction.

## III. DISCUSSION

The sensor is turned on and placed inside the magnetic shield. We program the sensor driver to output not only the field magnitude reading but also the signals for the field angle measurement. The direction of the magnetic field is then changed while its magnitude is kept the same at 50 µT. The signal V as a function of set polar angle, calculated from the currents going through the two Helmholtz coils, is shown as black dots in Figure 3 (a). A fit of Equation (1), using $C_0$, $C_1$ and $C_2$ as the fitting parameters, to the data set is also plotted as the red curve. As can be seen, the experimental data agrees well with the theoretical model. The fitting parameters are saved for the field angle measurement later on. We recorded the probe DC transmission signal V for 10 minutes at $\theta=45°$ and calculated its Allan Deviation [29]. The result is shown in Figure 3 (b). With an integration time of over 100 ms, the deviation of V drops below 0.3 mV. Combined with the slope of the curve in Figure 3 (a) which is about 0.05 degree/mV around $\theta=45°$, the most sensitive region of the angle measurement should have a sensitivity of better than 0.02° at the 10 Hz output rate. The measurement sensitivity decreases away from $\theta=45°$. Around $\theta=85°$, the sensitivity is worse than 0.2° with 100 ms measurement time. We repeat the same experiment with B = 20µT and 90µT. As expected, the signal V as a function of polar angle does not depend on the field magnitude.

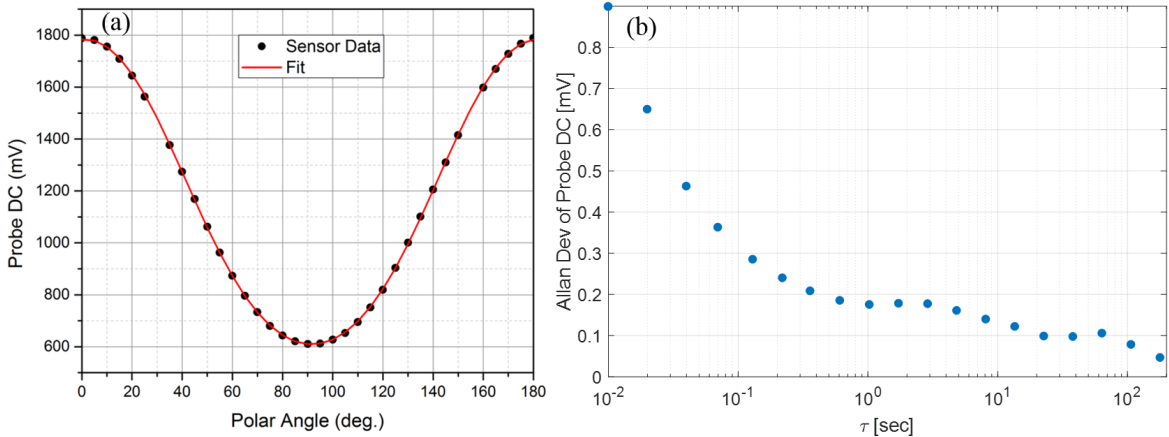

Figure 3 (a) Probe DC signal as a function of set polar angle (black dots). The red curve is a fit of Equation (1) to the data. (b) Allan deviation of the probe DC transmission signal V at $\theta=45°$.

The bandwidth of the polar angle measurement based on the probe DC transmission is also investigated. We set the polar angle to be 60°. A small oscillating field is added in the x-direction, perpendicular to the probe. As a result, the total polar angle is changed by a couple of degrees at the same frequency as the oscillating field. We fix the oscillating field amplitude and change its frequency. The amplitude of the measured oscillating V is recorded as a function of the driving frequency. The gain of the signal at a certain frequency is calculated as the normalized amplitude relative to that at 1 Hz. The resulting bandwidth data is plotted in Figure 4. As can be seen, the 3-dB bandwidth is well above 50 Hz. We did not test frequencies beyond 50 Hz as the maximum sample rate for signal V is 100 Hz due to technical limitations.

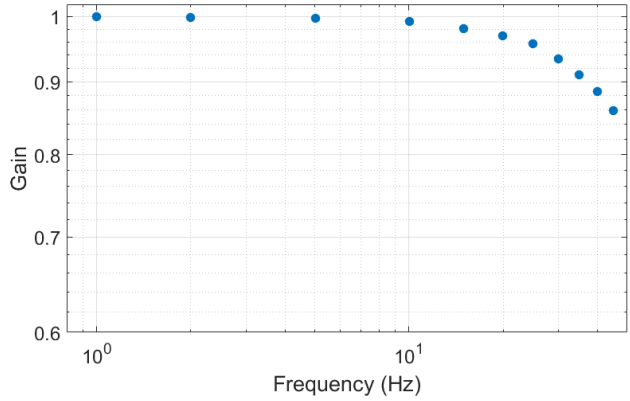

Figure 4 Bandwidth of the probe DC transmission based polar angle measurement.

To implement the light-shift based field angle measurement, a small 3 kHz modulation signal is generated by the sensor driver and applied to the probe laser current. The modulation amplitude is small enough to not affect the magnetometer performance. Inside the sensor driver, the scalar field reading is demodulated at 3 kHz with a bandwidth of 36 Hz. The demodulated signal, which should be equal to $B_{1k}cos\theta$, is included in the sensor output. We first measure the demodulated signal as a function of the set polar angle. The data is shown in Figure 5 (a) as black dots. A fit of a cosine function to the data is plotted as the red curve. The fitting yields the parameter $B_{1k}$. As shown in the figure, the data is well represented by a cosine curve, especially around $\theta=90°$. Far away from $\theta=90°$, the increased magnetometer noise and the imperfection in the probe light polarization can lead to reduced performance of this angle measurement method. The Allan deviation is also measured and shown in Figure 5 (b). With an integration time of 100 ms, the noise in the demodulated signal is about 40 pT. The slope of the curve in Figure 5 (a) is about 0.05 degree/pT around $\theta=90°$, which yields an angle measurement sensitivity of 2° at 10 Hz output rate. If the modulation frequency can be lowered to 300 Hz, close to the natural bandwidth of the magnetometer, the noise in the demodulated signal can be greatly reduced. However, the commercially available magnetometer has a 1 kHz sample rate for the magnetic field reading. Any extra signal below 1 kHz can potentially interfere with the normal magnetometer operation and therefore is prohibited.

The final $\theta$ is determined by the combination of these two methods. We first use the probe DC transmission signal $V$ to calculate $\theta$ from 0 to 90° and then rely on the supplementary method to determine whether the actual $\theta$ is <90° or >90°. When $V$ is within 5 mV from the minimum value ($\theta\approx90°$), $\theta$ is determined entirely from the supplementary method.

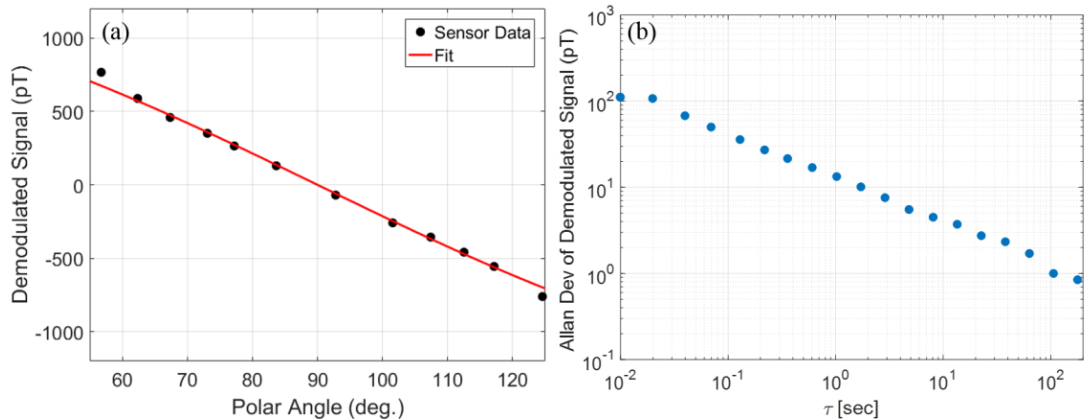

Figure 5 (a) Demodulated signal as a function of polar angle (black dots). The red curve is a fit of a cosine function to the data. (b) Allan deviation of the demodulated signal $V$ at $\theta=90°$.

Next we test the complete field angle measurement in an unshielded environment. We use the Geometrics MFAM module, which has two sensors connected to the same driver board and automatically synchronized. A testing facility is also set up at a location where the background magnetic field (Earth's field) is minimally disturbed. The two sensors are mounted to a rotation table as shown in Figure 6. The optical axis of sensor 1 is aligned with the rotation axis of the table and the sensor 2 optical axis lies parallel to the rotation plane. The rotation angle of the table is recorded by an optical encoder with a resolution of 0.35°. The rotation plane can also be adjusted between -30° and 70° with respect to the magnetic field (0º as shown in Figure 6).

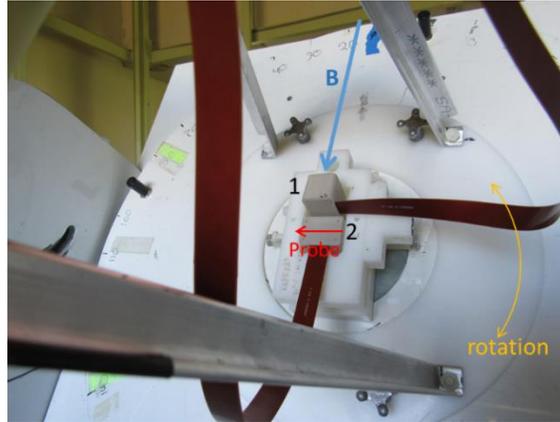

Figure 6 Testing setup for the field angle measurement in an unshielded environment.

To test the polar angle measurement, the table is aligned such that its rotation axis is perpendicular to the background magnetic field. In this configuration, the rotation angle of the table, given by the optical encoder, is the set polar angle of sensor 2. We record both $\theta$ and the optical encoder output and plot $\theta$ as a function of the rotation angle in Figure 7 (a). The measurement time for each point is 100 ms. The measurement error, defined as the difference between $\theta$ and the set polar angle, is also plotted in Figure 7 (b). As can be seen, for polar angles from 20° to 160°, the measurement produces fairly accurate results except in the vicinity of $\theta=90°$ where only the supplementary method is used. The noise in the error plot is mainly caused by the resolution of the optical encoder and not by the angle measurement.

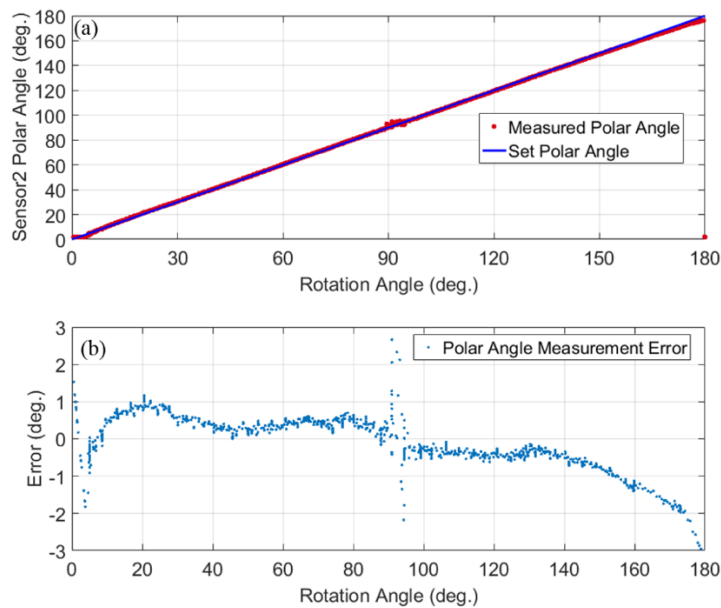

Figure 7 (a) Measured polar angle vs set polar angle. (b) Measurement error.

In the coordinate system defined by the two sensors with sensor 1 probe along z-direction and sensor 2 probe along x-direction, according to Equation (2) the azimuthal angle φ of the magnetic field can also be measured. We first tilt the rotation plane of the table by 40° with respect to the magnetic field so that its polar angle $\theta=130°$ in the coordinate system. We then rotate the table and record both the angle measurement results from the two sensors ($\theta_1, \theta_2$) and the optical encoder output. $\theta_1$ and $\theta_2$ as a function of the table rotation are shown in Figure 8 (a). As can be seen, $\theta_1$ does not change while $\theta_2$ varies between 45° and 150° during the table rotation from 0° to 180°. In the coordinate system, $\theta=\theta_1$ and φ can be calculated according to Equation (2) and should be the same as the optical encoder output. The result is shown in Figure 8 (b). As expected, φ as a function of the table rotation is linear from 0° to 180°.

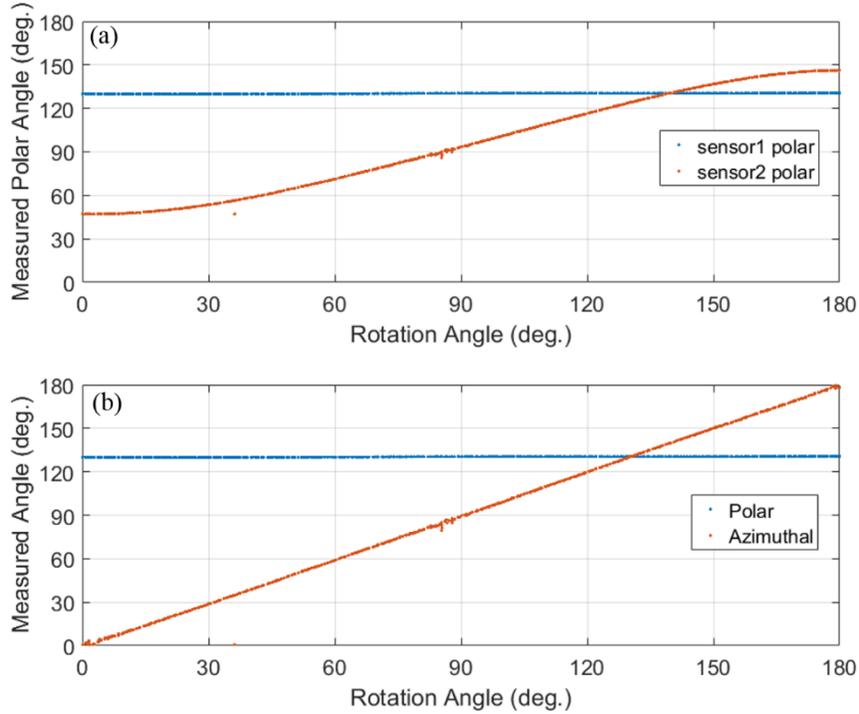

Figure 8 (a) Polar angle from both sensors as a function of the table rotation while the rotation plane has a 40° angle with respect to the magnetic field. (b) In the coordinate systems defined by the two sensors, both the polar angle and the azimuthal angle as a function of the table rotation.

The angle measurement accuracy is calibrated over a broad range of θ and φ. As discussed previously, the polar angle is set by adjusting the table rotation plane with respect to the background magnetic field with an accuracy of about 1°. At a set polar angle, the table is rotated and the set azimuthal angle is given by the optical encoder. We compare the measured angle with the set angle and define the total angle measurement error as $\sqrt{(\Delta\theta)^2+(\Delta\varphi)^2}$. For set polar angles between 60° and 160° in steps of 10° and set azimuthal angles from 0° to 180° in steps of 5°, the total angle measurement error is plotted on a color scale from 0° to 6° in Figure 9. Errors larger than 6° are shown in bright yellow. As can be seen, the angle measurement method can produce results with better than 3° accuracy for most sensor orientations. Approaching φ=0 or 180°, sensor 2 is close to its dead zone (optical axis aligned with the magnetic field). As shown in Figure 7 (b), inside or close to the dead zone, the polar angle measurement has large errors because the probe DC transmission $V$ is plateaued while the supplementary light-shift method either does not work or has huge noises. Note that a single set of fitting parameters ($C_0, C_1, C_2, B_{1k}$) for each sensor is used for the results shown in Figure 9.

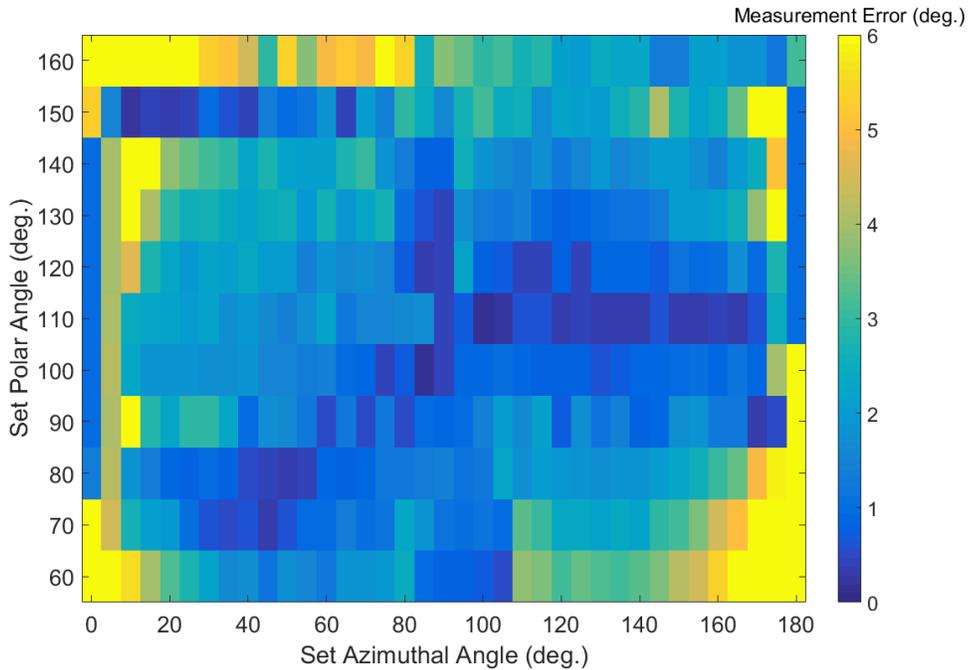

Figure 9 Total angle measurement error at different set angles.

## IV. CONCLUSION

In conclusion, we have demonstrated the vector field measurement using scalar atomic magnetometers. Away from the dead zones and 90° polar angle, a single sensor can produce polar angle measurements with better than 1° in accuracy and 0.05° in noise at 10 Hz sample rate. With two orthogonal sensors, the azimuthal angle of the field can also be measured with an accuracy of better than 3° for most sensor orientations. The measurement accuracy is expected to be greatly improved if we can lower the frequency of the probe current modulation to less than 300 Hz. This is possible with customized sensor drivers. Previous vector magnetometer research has not been able to address the measurement accuracy issue due to the large size of bench-top magnetometers. The best angle measurement sensitivity achieved in a finite magnetic field is about 0.003º with a measurement time of 100 ms [22], although using a much bigger cell. In practical applications, the biggest challenge facing the angle measurement using MFAM sensors is the stability of the operating temperature of the vapor cell. We do find that the ambient temperature change affects the probe beam transmission, which indicates that the cell operating temperature is coupled to the environment. To solve this problem, either the cell temperature stabilization needs an improvement or the sensor head has to be placed inside a thermally-isolated package. Another issue is the aging effect of the laser, causing the light power to change. In general, the laser aging is a long term effect. Regular laser power calibration should mitigate this problem.

## ACKNOWLEDGMENTS

This work is supported by the Strategic Environmental Research and Development Program (SERDP) under project number MR-2646.

## DATA AVAILABILITY

The data that support the findings of this study are available from the corresponding author upon reasonable request.